\documentclass[12pt]{report}
\usepackage{amsmath, amssymb, graphics}
\newcommand{\mathsym}[1]{{}}

\newcommand{\be}{\begin{equation}}
\newcommand{\ee}{\end{equation}}
\newcommand{\bc}{\begin{center}}
\newcommand{\ec}{\end{center}}
\newcommand{\bea}{\begin{eqnarray}}
\newcommand{\eea}{\end{eqnarray}}

\newcommand{\pr}{\prime}

\newcommand{\nn}{\nonumber}
\newcommand{\rf}{\ref}

\newcommand{\noi}{\noindent}

\newcommand{\al}{\alpha}

\newcommand{\lb}{\label}

\newcommand{\qu}{\quad}

\usepackage{amsmath, amssymb, graphics}

\begin{document}
\bc
{\large{\bf{Symmetry Analysis of 2+1 dimensional Burgers equation with variable damping}}}

D. Pandiaraja$^1$ and B. Mayil Vaganan$^2$

$^1$Department of Mathematics, Thiagarajar College, Madurai-625009, India\\
$^2$Department of Applied Mathematics and Statistics, 
Madurai Kamaraj University, Madurai-625021, India\\
\ec

\noi {\bf Abstract}
The symmetry classification of the two dimensional Burgers equation with variable coefficient is considered. Symmetry algebra is found and a classification of its subalgebras, up to conjugacy, is obtained. Similarity reductions are performed for each class.


{\bf AMS Classification Numbers} 22E60, 27E70, 34A05, 35G20.

{\bf 1.Introduction}
\setcounter{chapter}{1}
\setcounter{equation}{0}

Gandarias [5] has studied Type-II hidden symmetries   of the two dimensional Burgers equation 
\be
u_t+uu_x - u_{xx} - u_{yy} =0. \lb{I1}
\ee
In [4], applications of  (\ref{I1}) have been discussed. 

In this paper we provide a detailed  symmetry analysis of  the two dimensional Burgers equation with variable damping, viz.,
\be
u_t+uu_x + \alpha(t) u - u_{xx} - u_{yy} =0. \lb{I2}
\ee

One of the significant application of Lie symmetry groups to differential equations is to achieve a complete classification of its symmetry reductions. The symmetry properties and reductions of  certain differential equations   have been recently investigated (See references 10-12 in [1]). 
As described in [8], the classification of group invariant solutions requires a classification of subalgebras of the symmetry algebra into conjugacy classes under the adjoint action of the symmetry group.  

This paper is organized as follows. In  section 2, we perform a symmetry classification of  (\ref{I2}) and a classification of one-dimensional, two-dimensional, three-dimensional subalgebras of the symmetry algebra. In  section 3, we tabulate the reductions of  (\ref{I2}) under one-dimensional, two-dimensional, three-dimensional subalgebras. We summarize the results in  section 4.

{\bf 2. \ The symmetry albegra and classification of subalgebras }
\setcounter{chapter}{2}
\setcounter{equation}{0}

Equation (\rf{I2}) is assumed to be invariant under Lie group of infinitesimal transformations (Olver [7], Blumen and Kumei [3])
\be
x_i^{*} = x_i + \epsilon \xi_i(x, y, t, u) + O(\epsilon^2),\qu i=1,2,3,4, \lb{I3a}
\ee
where $\xi_1 =\xi, \xi_2 = \eta, \xi_3 =\tau, \xi_4=\phi$. Then the fourth prolongation $pr^{(4)}V$ of  
\be
V=\tau (x,y,t;u)\ \partial_t + \xi (x,y,t;u)\ \partial_x+\eta(x,y,t;u)\ \partial_y +\phi(x,y,t;u)\ \partial_u,
\ee
 must satisfy
\be
pr^{(4)}V\Omega(x,y,t;u)|_{\Omega (x,y,t;u)=0}=0,\lb{I4}
\ee 
where $\Omega$ is the RHS of (\ref{I2}).
The following system of 12 detemining equations are obtained from (\ref{I4}) (See [6]).
\bea
(\xi_1)_u = 0, \nn \\
(\xi_2)_u = 0, \nn \\
(\xi_3)_u = 0, \nn \\
(\phi_1)_{u,u} = 0, \nn \\
(\xi_3)_y = 0, \nn \\
(\xi_3)_x = 0, \nn \\
-(\xi_2)_t - u (\xi_2)_x + (\xi_2)_{x,x} + (\xi_2)_{y,y} -2 (\phi_1)_{y,u} = 0, \nn\\
\phi_1 - (\xi_1)_t - u(\xi_1)_x  + 2u(\xi_2)_y+(\xi_1)_{x,x}+(\xi_1)_{y,y} -2(\phi_1)_{x,u}= 0, \nn\\
u \alpha_t \xi_3 +\alpha \phi_1 + 2u \alpha(\xi_2)_y + (\phi_1)_t -u\alpha (\phi_1)_u +u (\phi_1)_x -(\phi_1)_{x,x}-(\phi_1)_{y,y}=0, \nn\\
(\xi_1)_y + (\xi_2)_x = 0, \nn\\
-(\xi_1)_x + (\xi_2)_y = 0, \nn\\
2(\xi_2)_y - (\xi_3)_t = 0. \nn
\eea 
Solving these determining equations we get 
\bea
\xi &=& \frac{q}{2}x + c_2 + c_1 \int{e^{-\int {\alpha}dt}dt}, \lb{a1}\\
\eta &=& \frac{q}{2}y + p, \lb{a2}\\
\tau &=& q t + r, \lb{a3}\\
\phi &=& -\frac{q}{2}u + c_1 e^{-\int {\alpha}dt},\lb{a4}
\eea
under the condition that 
\be
(q t + r )\alpha_t =0 \lb{I3}
\ee
Thus we have the following theorems:

Theorem 1. If $\alpha(t) = \alpha_0$, a real constant then symmetry algebra $L_c$ of (\ref{I2}) can be written as $L_c=V_1 + V_2 + V_3+V_4$ where $V_1 =-\frac{e^{-\alpha_0 t}}{\al_0} \partial_x + e^{-\alpha_0 t} \partial_u$, $V_2 = \partial_t$, $V_3 = \partial_y$ and $V_4 = \partial_x$ provided $q = 0, r \ne 0$

Table 1-
Commutator table for the Lie algebra $L_c$:
\bc
\begin{tabular}{|l|l|l|l|l|}
\hline
	 &$	V_1$	& $	V_2$	& $	V_3$	& $	V_4$ \\
\hline
$V_1$&	0&$\alpha_0 V_1$&$0$&$0$\\
\hline
$V_2$&$	-\alpha_0 V_1$&$0$&$0$&$0$\\
\hline
$V_3$&$0$&$0$&0&0\\
\hline
$V_4$&$0$&$0$&0&0\\
\hline
\end{tabular}
\ec

Theorem 2. If $\alpha(t)\ne 0$ is an arbitrary function of time t, then the general element of the symmetry algebra $L$ of (\ref{I2}) can be written as $L=V_1 + V_2 + V_3$ where $V_1 =\int{e^{\int {-\alpha}dt}dt}\partial_x + e^{\int {-\alpha}dt} \partial_u$, $V_2 = \partial_x$ and $V_3 = \partial_y$.
 
It should be noted here that the generators $V_1, V_2, V_3$ commute.

The problem of enumeration of all subalgebras of a given finite-dimensional Lie algebra L is essential for the group analysis of differential equations. For this group Int L of the inner automorphisms can be found. Under the operation of an automorphism, every algebra transforms into a subalgebra of the same dimensionality.

During the construction of optimal systems for a given Lie algebra L a special role is played by the center Z. The point is that every vector $z\in Z$ is inveriant relative to an automorphism $A \in Int L$. Thus the central elements generate subalgebras (ideals), which cannot changed by any automorphisms. They can be included as a direct term in any subalgebra of the Lie algebra L. Thus if the  optimal systems are known for the factor algebra L/Z, then they can be considered as known for the entire Lie algebra L.  It is sfficient to describe methods for constructing optimal systems only for Lie algebras with null centers [8].

{\bf Case 1}. $\alpha$ is a constant.

We assume that $\alpha=\alpha_0$.  
The commutator table gives the following information about the structure of the Lie algebra $L_c$.
The Lie algebra $L_c$ can be written as $L_c = Z \oplus L_2$ where $Z = \left\{V_3, V_4\right\}$ is the center of the Lie algebra  and $L_2 = \left\{V_1,V_2\right\}$

 We shall proceed to construct the optimal system now. First the mapping ad(V) is calculated by the equation
$ad(V)<x> = [x,V],$ where
 V is the general vector, $V= V^1 V_1+ V^2 V_2 $ and $x \in L_2$ and it turns out to be 

ad(V)= \begin{math}\bordermatrix{& & \cr
          & \alpha_0 V^2  & -\alpha_0 V^1  \cr
          & 0  & 0 \cr 
          }\end{math}\\
The genaral automorphism of the group Int $L^2$ can find by construting one-parameter groups of Automorphisms $A_i (t)$ for every basis vector $V_1,V_2$

To find the group $A_1(t)$ correspondingto $V_1$, it is neccessary to solve the equation $\partial_tx ^{'}=[x^{'},V]$, solving we get the matrix of automorphism

$A_1(t)$= \begin{math}\bordermatrix{& & \cr
          & 1 &  -\alpha_0 t  \cr
          & 0  & 1 \cr 
          }\end{math}\\

To find the group $A_2(t)$ correspondingto $V_2$, it is neccessary to solve the equation $\partial_tx ^{'}=[x^{'},V]$, solving we get the matrix of automorphism

$A_2(t)$= \begin{math}\bordermatrix{& & \cr
          & e^{\alpha_0 t} & 0  \cr
          & 0  & 1 \cr 
          }\end{math}\\

Here it is convenient to assume $t = \frac{1}{\alpha_0}a$ in $A_1(t)$ and $t = \frac{1}{\alpha_0}ln b$ in $A_2(t)$

\be
A = A_1(\frac{1}{\alpha_0}a) o A_2(\frac{1}{\alpha_0})ln b
\ee

The genaral automorphism $A \in Int L^2$ is obtained and it depends on two parameters, $a, b$:

A = \begin{math}\bordermatrix{& & \cr
          & b &  - a\cr
          & 0  & 1 \cr 
          }\end{math}\\
Now, it is possible to find the components of the vector  $x^{\pr} = A(x)$ in the bais $\left\{V_1,V_2\right\}$:
\be
{x^{\pr}}^1 = b x^{1} - a x^2,\quad 
{x^{\pr}}^2 = x^2 \lb{ab1}
\ee

The decomposition of the space $L^2$, on the classes of similar vectors, can be found using the invariance property of the killing form.

The killing form is 
\be
\frac{1}{\alpha_0^2}K(x,x) = (x^2)^2,
\ee
and is invariant of the group $Int L^2$. Since $K(x,y) \ne 0$, it is not degenerate and the algebra $L^2$ is semi simple. The vectors are separated into two disjoint classes:

1) $K(x,x) > 0$,\quad  2) $K(x,x)  =0$ 

A representative of class (1) is ${V_2}$ and a representative of class (2) is ${V_1}$. 
Hence $\Theta_1 = \left\{V_1,V_2 \right\}$.
Thus the sub algebras of the symmetry algebra $L_c$ is given in the following table:

Table 3
Classification of subalgebras of symmetry algebra $L_c$\\

\hrule									
Dimension \qu\qu\qu\qu\qu\qu\qu\qu\qu\qu\qu\qu\qu\qu\qu Subalgebra\\
\hrule
1-dimensional subalgebra \qu\qu\qu\qu\qu\qu\qu\qu $L_1 = ( aV_3 + bV_4 + V_1 )$
 
\qu\qu\qu\qu\qu\qu\qu\qu\qu\qu\qu\qu\qu\qu\qu\qu\qu\qu\qu $L_2 = ( aV_3 + bV_4 + V_2 )$
                               
\qu\qu\qu\qu\qu\qu\qu\qu\qu\qu\qu\qu\qu\qu\qu\qu\qu\qu\qu $L_3 = ( aV_3 + bV_4 )$

2-dimensional subalgebra \qu\qu\qu\qu\qu\qu\qu\qu $L_4 = ( aV_3 + bV_4 , V_1 )$
 
\qu\qu\qu\qu\qu\qu\qu\qu\qu\qu\qu\qu\qu\qu\qu\qu\qu\qu\qu $L_5 = ( aV_3 + bV_4 , V_2 )$
                               
\qu\qu\qu\qu\qu\qu\qu\qu\qu\qu\qu\qu\qu\qu\qu\qu\qu\qu\qu $L_6 = ( V_3 , V_4 )$

3-dimensional subalgebra \qu\qu\qu\qu\qu\qu\qu\qu $L_7 = ( aV_3 + bV_4 , V_1 , V_2 )$\\

\hrule

{\bf Case 2}. $\alpha$ is arbitrary

The commutator table gives the following information about the structure of the Lie algebra $L$.
The Lie algebra $L$ itself a center hence a maximal ideal. Thus the subalgebras of the symmetry algebra can be classified as in table 4. 

Table 4
Classification of subalgebras of symmetry algebra $L$\\

\hrule									
Dimension \qu\qu\qu\qu\qu\qu\qu\qu\qu\qu\qu\qu\qu\qu\qu Subalgebra \\
\hrule
1-dimensional subalgebra \qu\qu\qu\qu\qu\qu\qu\qu$L_1 = ( aV_1 + bV_2 + cV_3 )$
 
2-dimensional subalgebra \qu\qu\qu\qu\qu\qu\qu\qu$L_2 = ( V_1,V_2)$
 
\qu\qu\qu\qu\qu\qu\qu\qu\qu\qu\qu\qu\qu\qu\qu\qu\qu\qu\qu$L_3 = ( V_1 , V_3 )$
                               
\qu\qu\qu\qu\qu\qu\qu\qu\qu\qu\qu\qu\qu\qu\qu\qu\qu\qu\qu$L_4 = ( V_2 , V_3 )$

3-dimensional subalgebra \qu\qu\qu\qu\qu\qu\qu\qu$L_5 = ( V_1 , V_2 , V_3 )$ \\

\hrule

{\bf 3. Symmetry reduction for  (\ref{I2})}
\setcounter{chapter}{3}
\setcounter{equation}{0}

 For brevity, we present only two representative reductions that too for the case when  $\alpha$ is arbitrary below. The complete details of reduced equations for each subalgebra when $\alpha$ is either a constant or any function of $t$ are presented in table forms. When $\alpha$ is arbitrary, the reductions to PDEs and then to ODEs are given respectively in Appendix-A and Appendix-B. Similarly, when $\alpha$ is a constant, the respective reductions are given in Appendix-C and Appendix-D.

3.1  Reduction under one dimensional subalgebra $V_1 + V_2 + V_3$

The characteristic equation for $aV_1 + bV_2 + cV_3$ is 
\be
\frac{dx}{1+\int{e^{\int -\alpha(t)dt}}} = \frac{dy}{1} = \frac{dt}{0} = \frac{du}{e^{\int -\alpha(t)dt}}
\ee
Integrating the characteristic equation we get three similarity variables 
\be
\xi = x-y(1+{\int e^{-\int\alpha(t)dt}}) ,\eta = t, u=ye^{-\int\alpha(t)dt}+ F(\xi,\eta)
\ee
In terms of the similarity variables, (\ref{I2}) is reduced to 
\be
(1+(1+{\int e^{-\int\alpha(\eta)d\eta}})^2)F_{\xi\xi}-F_\eta - FF_\xi-\alpha(\eta) F=0
\ee
We give a complete table of reductions to PDEs for all one dimensional subalgebras in Appendix-A.

3.1  Reduction under two dimensional sub algebra $[V_2,V_3]$

Consider the algebra given by $V_2$ and $V_3$. Since $[V_2,V_3] =0$, we begin with $V_2 = \frac{\partial}{\partial x}$. The similarity variables for this generators are given by $\xi = y, \eta =t, u = F(\xi,\eta)$. Using these variables (\ref{I2}) reduces to a PDE 
\be
F_{\xi\xi}-F_\eta + \alpha(\eta) F=0. \label{Ij1}
\ee
In order to perform second reduction of the above equation, we firstly write $V_3$ in terms of new variables $\xi$, $\eta$ and $F(\xi,\eta)$ as 
$V_3 = $

\bc
{\bf Conclusions}
\ec

The symmetry classification of the two dimensional Burgers equations with constant and variable coefficients, viz.,
\bea
u_t+uu_x + \alpha_0 u &=& u_{xx} - u_{yy} =0,\\
u_t+uu_x + \alpha(t) u &=& u_{xx} - u_{yy} =0,
\eea
have been carried out. For, the symmetry algebras and their subalgebras, up to conjugacy, are enumerated. Successive   reductions to PDEs with two independent variables and then to ODEs of both second and first orders are performed for each subalgebra. As most of the reduced equations are of the standard forms or can be changed to canonical forms  through simple transformations we dispense with the problem of writing solutions to these equations. 

\bc
{\bf References} 
\ec
\begin{enumerate}
\item Azad H and  Mustafa M T \ 2007 \ {\it J. Math. Anal. Appl.}\ {\bf333}\ 1180-1188
\item Barbara Abraham Shrauner and Keshlan S Govinder \ 2006 \ {\it J. Nonlin. Math. Phys.)}\ {\bf13}\ 612-622
\item Bluman G W and Kumei S  \ {\it Symmetries and Differential Equations},\   Springer-Verlag, New York, 1989.
\item Edwards M P and Broadbridge P \ 1995 \ {\it Z. Angew. Math. Phys. }\ {\bf 46} 595 - 622
\item Gandarias M L \ 2008 \ {\it J. Math. Anal. Appl.}\ {\bf348}\ 752-759
\item Gerd Baumann\ {\it Symmetry Analysis of Differential Equations with Mathematica},\ Springer-Verlag, New York, 2000.
\item Olver P J   \ {\it Applications of Lie Groups to Differential
 Equations},\    Springer-Verlag, New York, 1986.
\item Ovsiannikov L V \ {\it Group analysis of Differential Equations}, Academic Press, New York, 1982
\end{enumerate}

{\bf Appendix-A}

\bc
\begin{tabular}{|l|l|}
\hline
\hline
$Subalgebra$&$Reduced \quad equation$ \\
\hline
\hline
$V_1$&$F_{\xi\xi}-F_\eta - \alpha(\eta) F=0$\\
\hline
$V_2$&$F_{\xi\xi}-F_\eta - \alpha(\eta) F=0$\\
\hline
$V_3$&$F_{\xi\xi}-F_\eta -FF_\xi- \alpha(\eta) F=0$\\
\hline
$V_1+V_2$&$F_{\xi\xi}-F_\eta - \alpha(\eta) F=0$\\
\hline
$V_2+V_3$&$2F_{\xi\xi}-F_\eta - FF_\xi-\alpha(\eta) F=0$\\
\hline
$V_1+V_3$&$(1+({\int e^{-\int\alpha(\eta)d\eta}})^2)F_{\xi\xi}-F_\eta - FF_\xi-\alpha(\eta) F=0$\\
\hline
$V_1+V_2+V_3$&$(1+(1+{\int e^{-\int\alpha(\eta)d\eta}})^2)F_{\xi\xi}-F_\eta - FF_\xi-\alpha(\eta) F=0$\\
\hline
\end{tabular}
\ec
{\bf Appendix B}
\bc
\begin{tabular}{|l|l|}
\hline
\hline
$Subalgebra$&$Reduced \quad equation$ \\
\hline
\hline
$[V_1,V_2]$&$ W_r + \alpha(r)W=0$\\
\hline
$[V_1,V_3]$&$W_r + \alpha(r)W=0$\\
\hline
$[V_2,V_3]$&$W_r + \alpha(r)W=0 $\\
\hline
$[V_1,V_2,V_3]$&$ W_r + \alpha(r)W=0 $\\
\hline
\end{tabular}
\ec

{\bf Appendix C}
\bc
\begin{tabular}{|l|l|l|}
\hline
\hline
$Subalgebra$&$Condition$&$Reduced \quad equation$ \\
\hline
\hline
$a V_3 + b V_4 + V_1$&$a \ne 0, b \ne 0$&$(a^2 +(b+e^{-\alpha_0 \eta})^2)F_{\xi\xi}-F F_\xi - F_\eta-  \alpha_0 F=0$\\
\hline
$$&$a =0 , b\ne 0$&$ F_{\xi\xi} - F_\eta-  \alpha_0 F=0$\\
\hline
$$&$a \ne 0,b=0$&$(a^2+\frac{e^{-2\alpha_0\eta}}{\alpha_0^2})F_{\xi\xi} - FF_\xi - F_\eta -\alpha_0 F=0$\\
\hline
$$&$a =0,b=0$&$F_{\xi\xi}-F_\eta =0$\\
\hline
$a V_3 + b V_4 + V_2$&$a \ne 0, b \ne 0$&$ F_{\xi\xi}+F_{\eta\eta} - a FF_\xi + bF_\xi+aF_\eta-\alpha_0 F=0$\\
\hline
$$&$a =0 , b\ne 0$&$F_{\xi\xi}+F_{\eta\eta} - FF_\xi + bF_\xi-\alpha_0 F=0$\\
\hline
$$&$a \ne 0,b=0$&$F_{\xi\xi}+F_{\eta\eta} - FF_\xi + aF_\eta-\alpha_0 F=0$\\
\hline
$$&$a=0,b=0 $&$F_{\xi\xi}+F_{\eta\eta} - FF_\xi -\alpha_0 F=0 $ \\
\hline
$a V_3 + b V_4$&$a \ne 0, b \ne 0 $&$ (a^2 + b^2) F_{\xi\xi} - a FF_\xi - F_\eta-\alpha_0 F=0$\\
\hline
$$&$a = 0, b \ne 0$&$F_{\xi\xi} -FF_\xi - F_\eta-\alpha_0 F=0 $ \\
\hline
$$&$a \ne 0, b =0$&$F_{\xi\xi}  - F_\eta-\alpha_0 F=0 $ \\
\hline
\end{tabular}
\ec

{\bf Appendix D}

\bc
\begin{tabular}{|l|l|l|}
\hline
\hline
$Subalgebra$&$Condition$&$Reduced \quad equation$ \\
\hline
\hline
$[a V_3 + b V_4,V_1]$&$a \ne 0, b \ne 0 $&$W_r + \alpha_0 W=0 $\\
\hline
$$&$a = 0, b \ne 0 $&$ W_r + \alpha_0 W=0$\\
\hline
$$&$a \ne 0, b =0  $&$ W_r + \alpha_0 W=0$\\
\hline
$[a V_3 + b V_4,V_2]$&$a \ne 0, b \ne 0 $&$(a^2+b^2) W_{rr}-aWW_r- \alpha_0 W = 0 $\\
\hline
$$&$a = 0, b \ne 0 $&$ W_{rr}- WW_r  - \alpha_0 W = 0 $\\
\hline
$$&$a \ne 0, b = 0 $&$W_{rr} - \alpha_0 W = 0 $\\
\hline
$[V_3 , V_4]$&$ $&$W_r + \alpha_0 W =0$\\
\hline
$[a V_3 + b V_4,V_1,V_2]$&$a \ne 0, b \ne 0 $&$ W_r + \alpha_0 W =0$\\
\hline
\end{tabular}
\ec

\end{document}